\documentclass[journal]{IEEEtran}
\ifCLASSINFOpdf
   \usepackage[pdftex]{graphicx}
   \usepackage{setspace}
   \usepackage{amsmath,epsfig,amssymb,subfigure,bm,dsfont}
   \usepackage{multirow}
   \usepackage{epstopdf}
   \usepackage{tabu}
   \usepackage{cite}
   \usepackage{soul}
   \usepackage{color}
   \usepackage{balance}
   \usepackage[citecolor=blue, colorlinks]{hyperref}
\else
\fi
\begin{document}
\title{Rank-Enhanced Low-Dimensional Convolution Set for Hyperspectral Image Denoising}


\author{Jinhui Hou, Zhiyu Zhu, Hui Liu, Junhui Hou,~\emph{Senior Member, IEEE}%,~% <-this % stops a space
%Huanqiang Zeng,~\emph{Senior Member, IEEE} %\\Jinjian Wu, \textit{Member, IEEE},
%and Jiantao Zhou,~\emph{Senior Member, IEEE}
\thanks{This work was supported by the Hong Kong Research Grants Council under
Grants CityU 11219019 and 11218121. The work of Hui Liu was supported by
Hong Kong University Grants Committee under the institutional development scheme research infrastructure grant UGC/IDS11/19. \textit{Corresponding author: Junhui Hou}}
% \emph{(Corresponding author: Junhui Hou)}
% \emph{$^\dag$ equal contribution, $^*$ corresponding author}
% <-this % stops a space
%\thanks{}
%\thanks{Copyright (c) 2015 IEEE. Personal use of this material is permitted. However, permission to use this material for any other purposes must be obtained from the IEEE by sending a request to pubs-permissions@ieee.org.}
% \thanks{$^\dag$ equal contribution, $^*$ corresponding author}
\thanks{J. Hou, Z. Zhu, and J. Hou are with the Department of Computer Science, City University of Hong Kong, Hong Kong (e-mail: jhhou3-c@my.cityu.edu.hk; zhiyuzhu2@my.cityu.edu.hk; jh.hou@cityu.edu.hk).}%
\thanks{H. Liu is with the School of Computing \& Information Sciences, Caritas Institute of Higher Education, Hong Kong. E-mail:hliu99-c@my.cityu.edu.hk}
%\thanks{H. Zeng is with the School of Engineering and School of Information Science and Engineering, Huaqiao University, Xiamen, China (e-mail: zeng0043@hqu.edu.cn).}
%\thanks{J. Wu is with the School of Artificial Intelligence, %Xidian University
%Xi'an, China (e-mail: jinjian.wu@mail.xidian.edu.cn).}
%\thanks{J. Zhou is with the Department of Computer and Information Science,
%University of Macau, Macau (e-mail: jtzhou@um.edu.mo).}
}
\maketitle

\begin{abstract}
This paper tackles the challenging %inverse
problem of hyperspectral (HS) image denoising.
%\st{To tackle such a challenging inverse restoration problem, existing deep learning-based methods designed complicated algorithms to optimize the networks so as to achieve remarkable performance.}
Unlike existing deep learning-based methods usually adopting complicated network architectures or empirically stacking off-the-shelf modules to pursue performance improvement, we focus on the efficient and effective feature extraction manner for capturing the high-dimensional characteristics of HS images.
%\st{propose a simple yet effective feature extraction module, namely rank-enhanced low-dimensional convolution set (Re-ConvSet), concentrating on exploring an efficient and effective feature extraction to capture the high-dimensional information of HS images.}
To be specific, %\st{we first unfold the high-dimensional filters into a 2D matrix to quantitatively estimate filter diversity, according to the rank of matrix.}
based on the theoretical analysis that increasing the rank of the matrix formed by the unfolded convolutional kernels can promote feature diversity,
%under the guidance of promoting the diversity of feature maps that is quantitatively measured, %\st{high-diversity and high-rank},
we propose rank-enhanced low-dimensional convolution set (Re-ConvSet), which separately performs 1-D convolution along the three dimensions of an HS image side-by-side, and then aggregates the resulting spatial-spectral embeddings via a learnable compression layer.
%\textcolor{magenta}{the resulting spatial-spectral embedding is further aggregated via learnable compression layer.}  %\st{formed of separate low-dimensional filters on different dimensions}.
Re-ConvSet not only learns the diverse spatial-spectral features of HS images, but also reduces the parameters and complexity of the network.
%\st{Finally, to construct our denoising framework, we embed the proposed module into a simple U-Net which is widely-used in the denoising task.}
We then incorporate Re-ConvSet into the widely-used U-Net architecture to construct an HS image denoising method.
%\st{Extensive experiments over commonly-used HS image denoising benchmarks demonstrate that}
Surprisingly, we observe such a concise framework %\st{Re-ConvSet can significantly promote the performance of classical learning-framework and}
outperforms the most recent method to a large extent in terms of quantitative metrics, visual results, and efficiency. %both quantitatively and visually. %i.e., \textcolor{magenta}{our method improves the PSNR value by XXX dB, while saving XXXX FLOPs}.
%built on \textcolor{magenta}{more advanced and complicated network architectures}. % with a large extent. in terms of restoration quality, parameters, and FLOPs.
We believe our work may shed light on deep learning-based HS image processing and analysis.
%The code will be publicly available at \url{XXXXXXXX}

\end{abstract}


% \fi
\IEEEpeerreviewmaketitle

\section{Introduction}

\IEEEPARstart
{O}{wing} Owing to the rich spectral and spatial information towards real-world scenes/objects, hyperspectral (HS) images have been adopted  %play a significant role
in numerous fields, such as military  \cite{shimoni2019hypersectral,jia2020status}, agriculture \cite{park2015hyperspectral,lu2020recent}, and marine monitoring \cite{Zhong2005Absorption,banerjee2020uav}.
Unfortunately, the captured HS images are inevitably corrupted by various noises due to the limitations of hardware and environment effects (e.g., water absorption and terrible atmosphere), which may severely degrade downstream applications.
% has serious restrictions on
% the development of
%\textcolor{blue}{the applications of HS images}.
% -based tasks
Thus, there is a gradually growing pursuit for HS image denoising algorithms. %\textcolor{blue}{[]}. %becomes a promising and essential direction.

\begin{figure*}[t]
\centering
\includegraphics[width=0.95\linewidth]{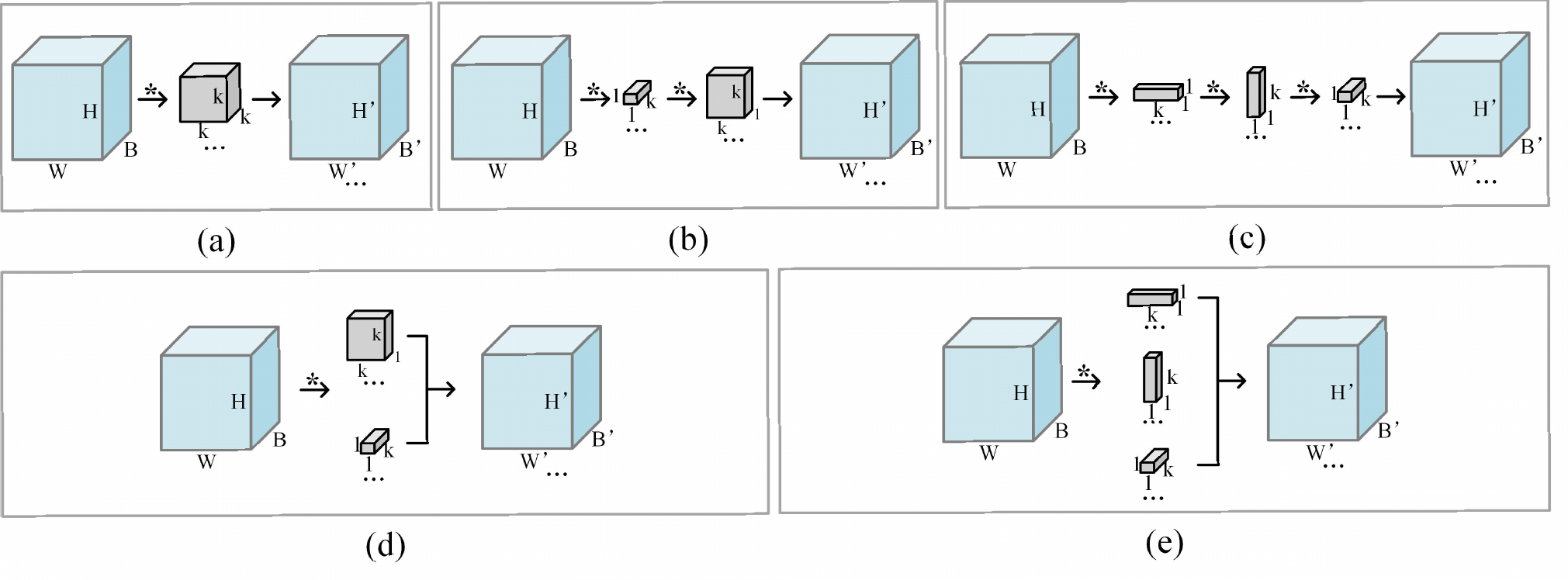}
\caption{Illustration of various potential feature extraction manners for HS images. %adopted  in existing deep learning-based HS denoising methods.
(a) 3-D convolution, (b) Sequential 1-D and 2-D convolution, (c) Sequential 1-D convolution, (d) 1-D + 2-D convolution, and (e) Proposed Re-ConvSet.}
\label{fig:low-rank-filters}
\end{figure*}

Over the past decades, many HS image denoising methods have been proposed \cite{Qian2013Hyperspectral,Fu2015Adaptive,Xie2016Multispectral,Cao2016Robust,He2020Non,Chang2019HSI,Dong2019Deep,Wei20213D,Cao2021Deep,Bodrito2021a}. The early works explicitly formulate HS image denoising as constrained optimization problems by employing prior knowledge, such as sparsity \cite{Fu2015Adaptive}, non-local similarity \cite{Qian2013Hyperspectral,Peng2014Decomposable}, total variation \cite{He2016Total,Wang2018Hyperspectral}, and low-rank tensors \cite{Xie2016Multispectral,Cao2016Robust,Chang2017Hyper,He2020Non}.
However, due to the limited representation ability, these methods are insufficient to model such an ill-posed inverse problem, making the quality of recovered HS images still unsatisfied.
%\textcolor{red}{Although the performance of these methods is comparable, they highly rely on well-designed handcrafted priors.} Besides, they are usually time-consuming due to the complex \st{need complex and costly} optimization processes.
Recently, owing to the powerful representational ability, deep learning-based HS image denoising methods have significantly improved the restoration quality \cite{Chang2019HSI,Yuan2019Hyperspectral,Dong2019Deep,Liu2019A,Zhang2019Hybrid,Wei20213D,Cao2021Deep,Bodrito2021a}.
%For the deep learning-based HS image denoising method, %\textcolor{blue}{the concerned issue lies in how to effectively extract intrinsic spatial-spectral embeddings from the high-dimensional HS images.
% the concerned issue lies in how to effectively extract the high-dimensional spatial-spectral information.
%Due to the high dimensionality, 3-D convolution is an intuitive choice for HS images processing.
%}
%For example,
Particularly,
%To tackle this issue,
most of existing deep learning-based methods extract spatial-spectral features of HS images by  %design the feature extraction module
by applying 2-D conventional filters with multiple channels on each cube slice separately \cite{Chang2019HSI}, 3-D conventional filters to simultaneously convolve in spatial and spectral domains \cite{Zhang2019Hybrid}, or a 2-D and 3-D conventional filters combined network \cite{Yuan2019Hyperspectral}. %See Section \ref{sec:Re} for more details.
However, these empirical designed feature extraction modules summarized in Fig. \ref{fig:low-rank-filters} are hard to provide quantitative instructions for network designing and may be not optimal, thus limiting performance.
% However, the 3-D convolution will introduce a significant increase in the trainable parameters, resulting in both high computational cost and the redundancy of networks.
%\textcolor{blue}{\textit{However,  those pioneering works only proposed qualitative analysis which is hard to give the quantitative instruction for the network designing. }}
Besides, most
%existing deep learning-based methods
of them tend to stack off-the-shelf modules to build complicated and large-capacity networks for pursuing high performance, such as multi-scale strategy \cite{Yuan2019Hyperspectral,Zhang2019Hybrid,Shi2021Hyperspectral}, attention mechanism \cite{Shi2021Hyperspectral}, and atrous convolution \cite{Liu2019A,Shi2021Hyperspectral}.

In contrast to existing works, %that usually adopt complicated network architectures, stack off-the-shelf deep modules, or empirically designed feature extraction layers for pursuing performance improvement,
we aim to embed the high-dimensional spatial-spectral information of HS images both efficiently and effectively via theoretical analysis. %Specifically, based on xxxxxx, we propose %a
Specifically, we first figure out the bottleneck that limits feature diversity by means of matrix rank analysis.
Under the guidance of the theoretical analysis, motivating us to propose Re-ConvSet, which separately performs 1-D convolution along the three dimensions of an HS image side-by-side, and then aggregates the resulting spatial-spectral embeddings via a compression layer. Such a manner improves the upper bound of the rank of the matrix formed by the convolution kernels invovled in Re-ConvSet to promote diverse features which are expected to be beneficial to image reconstruction. Meanwhile, Re-ConvSet reduces the number of network parameters.
%is composed of separate low-dimensional filters on different dimensions, to effectively obtain the informative spatial-spectral features of HS images.
Finally, we realize an \textit{efficient}, \textit{concise}, and \textit{compact} denoising method by incorporating
%the proposed feature extraction module into a simple
Re-ConvSet into the widely-used U-Net architecture.
Extensive experiments on both synthetic and real noisy HS images  demonstrate the significant superiority of the proposed denoising method over state-of-the-art ones. %Besides, we comprehensively evaluate, which xxxxx

In summary, the main contributions of this paper are three-fold:% mainly lie in following three aspects:
\begin{itemize}

\item %from the perspective of promoting the diversity of feature maps, we propose
an efficient and effective spatial-spectral feature extraction module based on theoretical analysis; %to capture the high-dimensional spatial-spectral information of HS images.

\item  comprehensive and quantitative investigations on various combinations of low-dimensional convolution that may shed light on deep learning-based HS image processing; and

\item  a \textit{concise} and \textit{compact} HS image denoisng method that reveals the importance of feature extraction and achieves the current state-of-the-art. %

\end{itemize}

The remainder of the paper is organized as follows. Section~\ref{sec:Re} reviews the existing related works. Section~\ref{sec:proposed} describes our proposed ReConv-Set in detail, followed by extensive experiments and analysis in Section~\ref{sec:experiments}. Finally, Section~\ref{sec:con} concludes this paper.

\section{Related Work}
\label{sec:Re}
%\textcolor{red}{[!!Please enrich this section to one page]}
\subsection{Optimization-based Methods}
 This kind of methods generally formulates the HS image denoising as complex optimization problems, relying on the well-designed handcrafted priors, such as non-local similarity \cite{Maggioni2013Nonlocal,Qian2013Hyperspectral,Peng2014Decomposable}, total variation \cite{He2016Total,Wang2018Hyperspectral} and low-rank priors \cite{Zhang2014Hyperspectral,Xie2016Multispectral,Cao2016Robust,Chang2017Hyper,Fan2017Hyperspectral,Chen2018Denoising,He2020Non}. To be specific, Qian \emph{et al.} \cite{Qian2013Hyperspectral} proposed a sparse representation-based method by introducing the nonlocal similarity and spectral-spatial structure of HS imagery. Peng \emph{et al.} \cite{Peng2014Decomposable} presented a novel tensor dictionary learning (TDL) model by taking the non-local similarity in space and global correlation in spectrum into account. He \emph{et al.} \cite{He2016Total} proposed a total variation regularized low-rank matrix factorization (LRTV) method, in which the nuclear norm, TV regularization, and $L_1$ norm were integrated.
 Owing to the consideration of modeling the intrinsic property of HS images, several nonlocal low-rank tensor-based models achieved remarkable denoising performance.
 Xie \emph{et al.} \cite{Xie2016Multispectral} proposed an intrinsic tensor sparsity regularization (ITSReg) model, in which the global correlation along the spectrum and nonlocal self-similarity across space were fully considered. Chang \emph{et al.} \cite{Chang2017Hyper} designed a hyper-Laplacian regularized unidirectional low-rank tensor recovery (LLRT) model to exploit the intrinsic structure correlation of HS image.  He \emph{et al.} \cite{He2020Non} provided a unified paradigm to fuse spatial non-local similarity and global spectral low-rank properties by utilizing the low-dimensional orthogonal basis. \\

%\textcolor{red}{Despite some traditional methods have achieved impressive performance, they highly rely on prior knowledge.
%and the denoising processes
%Besides, they are usually time-consuming.}

\subsection{Deep Learning-based Methods}
 %\st{Owing to the powerful representation ability of deep neural networks,}
 A considerable number of deep learning-based methods for HS image denoising have been presented, which improve the restoration quality of traditional optimization-based methods dramatically. For example,
 Chang \emph{et al.} \cite{Chang2019HSI} first tackled HS image denoising by a deep neural network, in which the learned 2-D filters with multiple channels were utilized.
 Yuan \emph{et al.} \cite{Yuan2019Hyperspectral} designed a novel spatial-spectral network with both 2D and 3D convolutional kernels to fully exploit spatial and spectral features.
 Liu \emph{et al.} \cite{Liu2019A} presented a 3-D atrous denoising convolution neural network (3DADCNN), in which 3-D kernels are integrated with the atrous convolution for enlarging the receptive fields in both the spatial and spectral dimensions simultaneously.
 Inspired by the separable 3-D spatial-temporal convolution\cite{Qiu2017Learning}, Dong \emph{et al.} \cite{Dong2019Deep} proposed a separable 3-D convolution network to explore spatial-spectral correlations by decomposing 3-D convolution into the concatenation of 2-D spatial convolution and 1-D spectral convolution.
 Zhang \emph{et al.} \cite{Zhang2019Hybrid} designed a multi-scale spatial-spectral convolutional network, in which the spatial gradient and spectral gradient are jointly incorporated.
 Maffei \emph{et al.} \cite{Maffei2020A} proposed a single network, taking a full HS image cube as input, to explore the spatial-spectral correlation.
 Wei \emph{et al.} \cite{Wei20213D} proposed a 3-D quasi-recurrent neural network (QRNN3D) to simultaneously explore the structural spatiospectral correlation and global correlation along spectral. Besides, an alternating directional structure was integrated to alleviate the spatiospectral dependence modeling.
 Shi \emph{et al.} \cite{Shi2021Hyperspectral} designed a 3-D attention denoising network (3-D-ADNet) with two parallel branches, including spatial branch with the position attention module and spectral branch with the channel attention module.
 Cao \emph{et al.} \cite{Cao2021Deep} designed a deep spatial-spectral global reasoning network (GRN) to explore the contextual information by combining the local and global spatial-spectral information of HS images.
 Rui \emph{et al.} \cite{Rui2021CVPR} presented a data-driven method to capture the general weighting principle of HS image denoising model.
 Bodrito \emph{et al.} \cite{Bodrito2021a} proposed a trainable spectral-spatial sparse coding (T3SC) model by employing sparse coding and deep learning.
% For ease of comparison, we show different feature extraction manners in Fig. \ref{fig:low-rank-filters}.

 \begin{figure}[!t]
\centering
\includegraphics[width=1\linewidth]{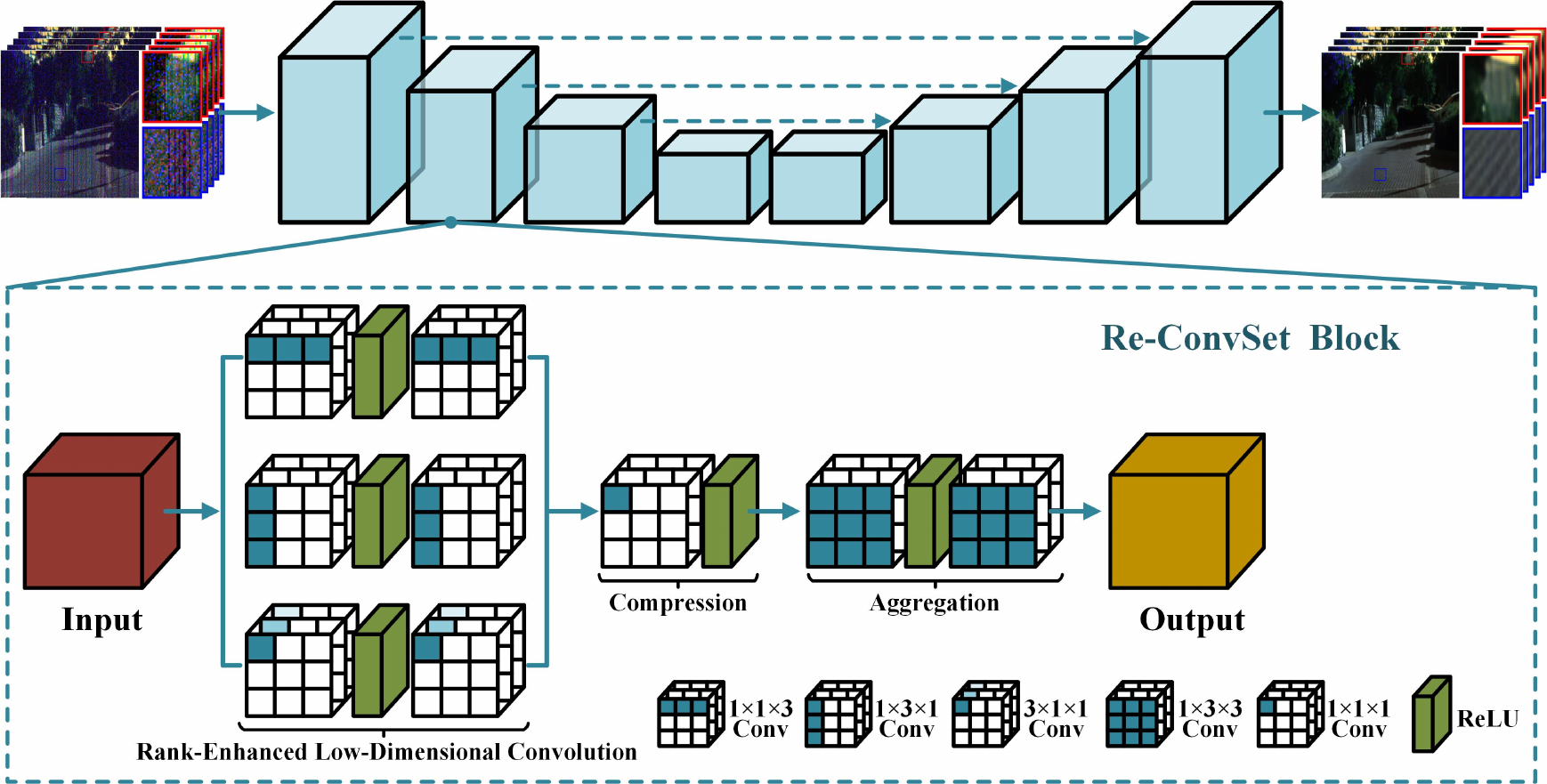}
\caption{Illustration of our HS image denoising framework, which is constructed by incorporating the proposed Re-ConvSet into a residual U-Net architecture.
%We refer the readers to \textit{Supplementary Material} for the detailed network configuration.
}
%\vspace{-0.4cm}
\label{fig:hsisr-framework}
\end{figure}

\section{Proposed Method}
\label{sec:proposed}

%In this section, we first formulate the problem for single HS image SR and present the motivation of the proposed method. Then, we introduce each module of the proposed framework in detail.

\subsection{Problem Statement}% Statement and Overview Motivation
\label{sec:Pfam}
Let $\mathcal{X}\in\mathbb{R}^{B\times H\times W}$ be a noisy HS image, and $\mathcal{Y}\in\mathbb{R}^{B\times H\times W}$ the corresponding noise-free one, where $H$ and $W$ are the spatial dimensions, and $B$ is the number of spectral bands. The degradation process of $\mathcal{Y}$ to $\mathcal{X}$ could be generally formulated as %\st{, we aim to recover a noise-free image \st{denoted as} $\mathbf{Y}$ $\in\mathbb{R}^{B\times H \times W}$.
% The corrupted process of $\mathbf{X}$ from $\mathbf{Y}$ can be linearly formulated as
%The additive noise model can be linearly formulated as}
\begin{equation}
%\small
\label{equ:1}
\mathcal{X} = \mathcal{Y} + \mathcal{N}_{z},
\end{equation}
where $\mathcal{N}_{z}\in \mathbb{R}^{B\times H \times W}$ denotes the additive noise. %random noise.
%consisting of the Gaussian noise, stripes, deadlines, impulse noise, or a mixed one of them.
Recovering $\mathcal{Y}$ from $\mathcal{X}$  is an  ill-posed inverse problem in high-dimensional space, making it very challenging. %\st{To tackle such an inverse reconstruction problem,}
Owing to the powerful representation ability and large capacity, recent deep CNN techniques %convolutional neural network %\st{is an intuitive choice, which}
have shown the great potential in addressing this problem
%\st{HS image processing}
\cite{Yuan2019Hyperspectral,Dong2019Deep,Zhang2019Hybrid,Wei20213D,Cao2021Deep}.
%\textcolor{cyan}{
Particularly, designing an efficient and effective learning backbone, i.e., the feature extraction layers, is one of the  most critical issues. %However, currently there are very limited knowledge giving us the theoretical instruction for such high-dimensional signal processing.
%}
% We argue that how to learn representative spatial-spectral features efficiently and effectively
% %\st{via well-designed feature extraction module}
% is critical issue.
%\st{one of the most critical issue for deep learning-based HS image denoising.} %\st{Due to the high dimensionality of HS image}

Intuitively, 3-D convolution is the most direct choice for capturing the high-dimensional spatial-spectral information of HS images. %the high-dimensional HS image feature extraction.
However, it leads to a significant increase in the parameter size, which may potentially cause over-fitting and consume huge computing resources. Actually, the increase in the number of parameters does not  bring about obvious performance improvement (see the results in Section \ref{subsubsec:abexp-filterrank}). %\st{To tackle this problem,}
Although some low-dimensional convolution-based feature extraction manners have been proposed,  %to perform HS feature extraction,
as reviewed in Section \ref{sec:Re}, %\st{to efficiently exploit the high-dimensional spatial-spectral information of HS images},
they were empirically designed on the basis of the HS data structure, making it hard to provide quantitative instructions. Besides, they may not be optimal, thus limiting performance. %See Sec.~\ref{sec:discussion} for more analysis towards low-dimensional feature extraction.
Different from existing manner, we propose an efficient and effective feature extraction manner named Re-ConvSet from the theoretical perspective of promoting feature diversity.
%\st{with the guidance of quantitative analysis}.

 We then incorporate the proposed Re-ConvSet into the widely-used U-Net \cite{ronneberger2015u} architecture for constructing an efficient and compact HS image denoising method, as illustrated in Fig.~\ref{fig:hsisr-framework}. We train the network by minimizing the $\ell_1$ distance between the recovered HS image $\mathcal{\widehat{Y}}$ and the corresponding noise-free HS image $\mathcal{Y}$:
\begin{equation}
%\small
\label{equ:l1loss}
\mathcal{L}_1(\mathcal{\widehat{Y}}, \mathcal{Y}) = \frac{1}{B\times HW}\left\|\mathcal{\widehat{Y}}- \mathcal{Y} \right\|_1.
\end{equation}

%In what follows, we will first ... followed by the . We finall
%detail the proposed Re-ConvSet.
\begin{figure}[!t]
\centering
\includegraphics[width=1\linewidth]{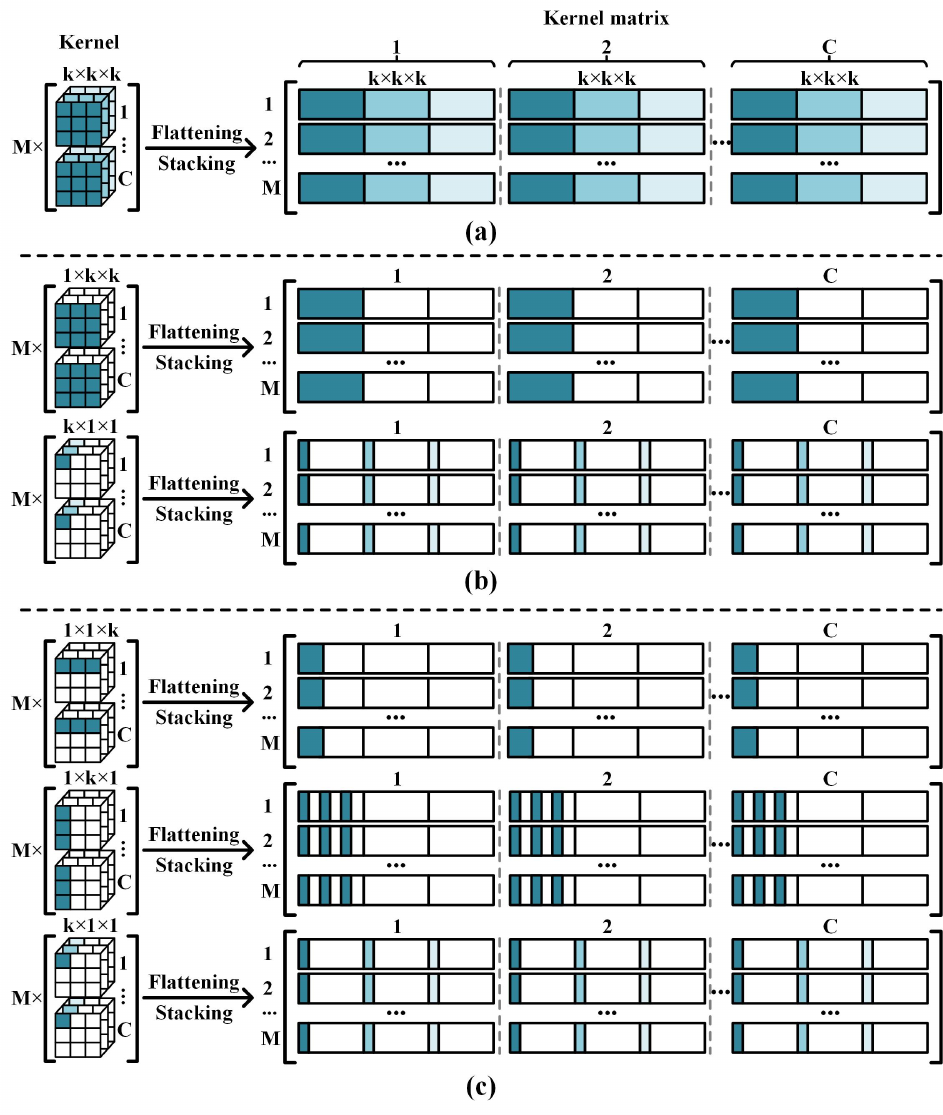}
\caption{Illustration of mapping the convolutional kernel into a matrix during the feature embedding process.
%For ease of demonstration, we set the kernel size $k$ to 3.
Specifically, the small cubes with colors stand for the kernel weights, and the blank ones represent zero. (a) 3-D convolutional kernel, (b) 1-D and 2-D convolutional kernels, and (c)  proposed Re-ConvSet. %\textcolor{red}{[Update this figure!!]}
}
\label{fig:filter-rank}
\end{figure}

\subsection{Feature Diversity Analysis} %? \st{Estimation}

Feature diversity can effectively reflect the information richness of feature maps that is positively correlated with network performance \cite{Ayinde2019Regularizing}.
%\st{With such a characteristic, the performance of network is vitally correlated with the feature maps with rich information \cite{}.}
Besides, some recent studies \cite{kim2018attention,Ayinde2019Regularizing,milbich2020diva,Zhong2021Selection} indicate that learning diverse feature maps in the context of deep CNNs show great capability in reducing overfitting and improving the generalization ability of networks.
%\st{Evidently, enhancing feature diversity during extracting features from high-dimensional HS images may bring immense benefits.}
Here, taking the 3-D convolutional layer as an example, we theoretically analyze the learned feature map with the rank of matrices and figure out the bottleneck limiting feature diversity, which further motivates us to design both efficient and effective feature representation module.

Let $\mathcal{A}\in\mathbb{R}^{M\times C\times k\times k\times k}$ denote a typical 3-D convolutional layer equipped with $M$ 3-D kernels  of size $k\times k\times k$, where  $C$ is the number of channels. When feeding an HS feature map $\mathcal{I}\in \mathbb{R}^{C\times B\times H\times W}$ into $\mathcal{A}$,  we can obtain an output feature volume  $\mathcal{F}\in \mathbb{R}^{M\times B'\times H'\times W'}$,  whose $m$-th feature map $\mathcal{F}_m \in \mathbb{R}^{B'\times H'\times W'}$ is obtained as %denotes the. i.e.,
%\st{by employing the mapping function}:
\begin{equation}
%\small
\label{equ:3dfilter}
% \mathcal{F}=\mathcal{A}*\mathcal{I}~~
\mathcal{F}_{m} = \sum^C_{c=1}\mathcal{A}^c_m\ast \mathcal{I}^c,~m=1,~2,~3,~\cdots, M,
\end{equation}
where $\ast$ is the convolution operator, and $\mathcal{A}_m$ stands for the $m$-th convolutional kernel %($m\in [1,2,...,M]$),
%which
consisting of a stack of 3-D kernels $\mathcal{A}^c_m \in \mathbb{R}^{k\times k\times k}$ ($1\leq c\leq C$).
By unfolding the high-dimensional tensors%\st{the filter $\mathcal{A}$ and input feature $\mathcal{I}$}
, we can equivalently re-write Eq. (\ref{equ:3dfilter}) in the form of 2-D matrix multiplication, i.e.,
\begin{equation}
\label{equ:convmatrix}
\mathbf{F} = \mathbf{A} \cdot \mathbf{I},
\end{equation}
where $\mathbf{F} \in \mathbb{R}^{M \times B'H'W'}$ is the matrix form of
%the output feature volume
$\mathcal{F}$, $\mathbf{A} \in \mathbb{R}^{M \times k^3C}$ is the kernel matrix generated by flattening each of the 3-D kernels as a row vector and then stacking all vertically, and $\mathbf{I} \in \mathbb{R}^{k^3C \times B'H'W'}$ denotes the matrix form of $\mathcal{I}$ generated by sliding the kernel across the features along the horizontal-spatial, vertical-spatial, and spectral directions.

%According to Eq. (\ref{equ:convmatrix}),
To quantitatively estimate the feature diversity, we analyze the rank of the feature matrix $\mathbf{F}$, which indicates the independence/freedom of elements of $\mathbf{F}$, and has demonstrated its effectiveness \cite{Lin2020HRank}.
%\st{been turned out to an effective measure of the information in matrix}.
Specifically, based on the property of matrix multiplication, %it can be seen from Eq. (\ref{equ:convmatrix}) that
from  Eq. (\ref{equ:convmatrix}) we have
\begin{equation}
\mathsf{Rank}(\mathbf{F}) \leq \mathsf{min}\{\mathsf{Rank}(\mathbf{A}), \mathsf{Rank}(\mathbf{I})\},
\end{equation}
where $\mathsf{Rank}(\cdot)$ returns the rank of a matrix.
%\st{$\mathsf{min}\{a,b\}$ represents the minimum of $a$ and $b$}.
Moreover, as
the values of $M$ and $C$ are usually comparable, we have $M \ll k^3C \ll B'H'W'$, resulting in %. Thus, we can conclude that %It denotes that the
$\mathsf{Rank}(\mathbf{F})\leq M$. In other words, the feature diversity is severely limited by the massive gap between the two dimensions of matrix $\mathbf{A}$.
Therefore, we can relive %our key problem lies in relieving
such an unbalanced matrix form of $\mathbf{A}$ to boost its rank upper bound, and likewise $\mathsf{Rank}(\mathbf{F})$, i.e., more elements of $\mathbf{F}$ are independent. %more fr produce more diverse features.
An intuitive way to boost the upper bound of $\mathsf{Rank}(\mathbf{A})$ is to increase $M$. However, it can be seen from Eq. (\ref{equ:3dfilter}) that the computational cost for a 3-D convolutional layer is $\mathcal{O}(CMk^3B'H'W')$, indicating that increasing $M$ directly will further lead to a significant increase of the parameter size and computational complexity.
%By quantitatively exploring the filter diversity,
Alternatively, we aim to improve the upper limit of
%Ultimately, motivated by enhancing feature diversity, our goal falls in increasing the rank of
$\mathsf{Rank}(\mathbf{A})$ without introducing extra computational burden to the network.
%Thus our key problem lies in how to alleviate the computational burden of the high-dimensional 3-D filters, while enhancing the filter diversity.

\subsection{Rank-Enhanced Low-Dimensional Convolution Set} %(Re-ConvSet)
%As shown in Fig. \ref{fig:filter-rank} (a),
Based on the above analysis, the problem boils down to
how to utilize the same or fewer elements shown in Fig. \ref{fig:filter-rank} (a) to form a kernel matrix with a  higher rank upper bound. To realize the goal, we can make the row-column sparse via filling these elements into different rows and columns of a larger zero matrix. %so as to achieve the higher rank.
Moreover, considering that the principal components of the 3-D kernel play a crucial role during the feature embedding process, we only employ the elements of $\mathbf{A}$ located at the principal components to fill the larger zero matrix, forming a new kernel matrix $\mathbf{A}_{rc} \in \mathbb{R}^{3M \times k^3C}$. %as shown in Fig. \ref{fig:filter-rank} (c).
Formally, we %construct %a rank-enhanced low-dimensional convolution set, named
can write the matrix form of Re-ConvSet as
%One critical issue in enhancing the filter diversity discussed above is how to achieve higher rank for the filter matrix by keeping the same or less elements.
%Fortunately, we empirically observe that the higher rank can be obtained by filling these elements into the diagonal of a larger zero matrix, which aims to increase the aspect ratio of the filter matrix.
%To keep the same or less parameters, some elements of filter matrix should be filled with zeros.
%Besides, considering that the principal components of the filter play a critical role during the feature embedding process, with the guidance of high-rank, we construct a rank-enhanced low-dimensional convolution set, named Re-ConvSet, by keeping these principal component elements in spatial-spectral domain (i.e., horizontal-spatial, vertical-spatial, and spectral directions) and filling them into a larger zero matrix to form a new filter matrix $\mathbf{A}_{rc} \in \mathbb{R}^{3M \times k^3C}$, as shown in Fig. \ref{fig:filter-rank} (c).
%we fill the elements of filter matrix outside the principal component with zeros.
%Hence, we construct a rank-enhanced low-dimensional convolution set in the spatial-spectral domain, which extracts representative spatial-spectral features of HS images via separately convolving on different dimensions, including horizontal-spatial, vertical-spatial, and spectral directions.
%The rank-enhanced low-dimensional convolution
%whose operation in matrix multiplication can be written as
\begin{equation}
\begin{split}
\label{equ:rankenchance-conv}
\mathbf{F}_{rc} = \mathbf{A}_{rc} \cdot \mathbf{I}~~~{\rm with}~~
\mathbf{A}_{rc} = [\mathbf{A}_{rc1};\mathbf{A}_{rc2};\mathbf{A}_{rc3}], %\in\mathbb{R}^{3M \times k^3C},
\end{split}
\end{equation}
where %$\mathbf{C}_{rc}\in \mathbb{R}^{M \times 3M}$ denotes the matrix of concatenation layer with $1\times 1\times 1$ convolutional filter.
$\mathbf{F}_{rc}\in \mathbb{R}^{3M \times B'H'W'}$ is the matrix form of the output feature volume extracted by our Re-ConvSet,
%$\mathbf{A}_{rc}\in \mathbb{R}^{3M \times k^3C}$ denotes the matrix of our Re-ConvSet.
and $\mathbf{A}_{rc1}$, $\mathbf{A}_{rc2}$, and $\mathbf{A}_{rc3} \in \mathbb{R}^{M \times k^3C}$ are the augmented matrix representations of low-dimensional kernels with zeros in  various spatial-spectral domains, corresponding to the kernels of  size $1\times 1\times k$, $1\times k\times 1$, and $k\times 1\times 1$, respectively.
%The computation process is shown in Fig. \ref{fig:low-rank-filters} (c).

As illustrated in Fig. \ref{fig:filter-rank} (c), $\mathbf{A}_{rc}$ is filled with many elements equal to zero, resulting in $7C$ valid columns. Due to $3M < 7C$ in practical implementations, we generally have $\mathsf{Rank}(\mathbf{A}_{rc})\leq 3M$, and thus $\mathsf{Rank}(\mathbf{F}_{rc})\leq 3M$.
%Considering that the $\mathbf{A}_{rc}$ is filled with a lot of zero elements, $\mathbf{A}_{rc}$ is actually $3M \times 7C$ matrix, where $3M < 7C$ in our settings.
%the rank of our Re-ConvSet matrix can be calculated as $\mathsf{Rank}(\mathbf{A}_{rc})\leq 3M$.
Compared with the original kernel matrix $\mathbf{A}$,
%the filter diversity is significantly improved in our Re-ConvSet matrix $\mathbf{A}_{rc}$.
our Re-ConvSet boosts the rank upper bound
%of corresponding kernel matrix
from $M$ to $3M$, thus potentially promoting more diverse features. %diversity of feature maps.
%during feature extraction process.
Besides, the zero elements in $\mathbf{A}_{rc}$ do not contribute to the network parameters.
Therefore, our Re-ConvSet not only learns the diverse spatial-spectral
features of HS image,
%can enhance the filter diversity to capture the high-dimensional information of HS images via different low-dimensional filters in spatial-spectral domain,
but also reduces the parameter size of the network.
%We also refer the readers to \textit{Supplementary Material} for the distribution of the singular values of the feature matrix, which indicates the independence of the elements more accurately.
%By far, our Re-ConvSet boosts the upper limit rank of the corresponding original 3-D filter matrix from $M$ to $3M$, enhancing the feature diversity during feature extraction process.

\begin{figure}[t]
  \centering
  \includegraphics[width=0.8\linewidth]{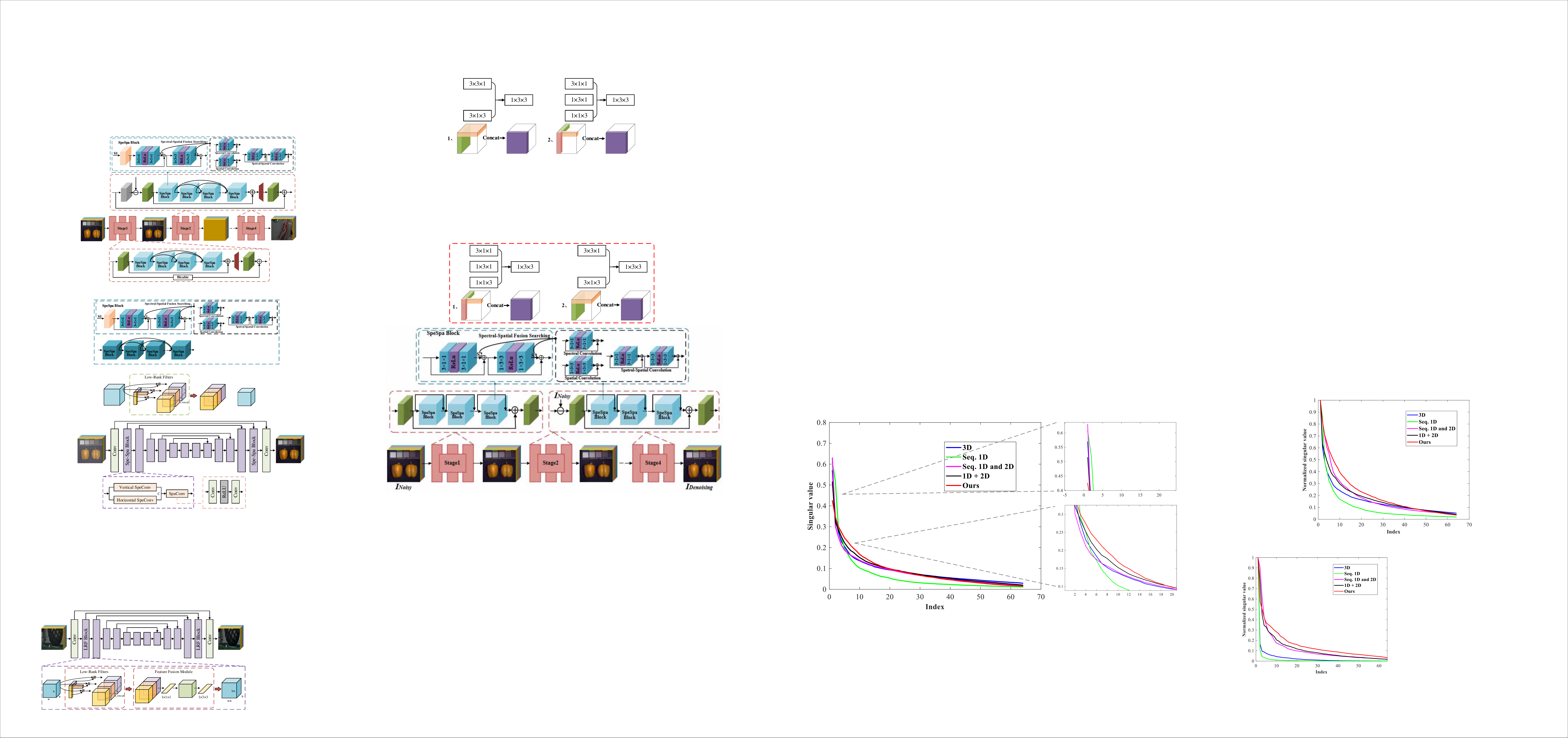}
   \caption{\footnotesize{Comparison of the singular value distributions of the feature maps extracted by different convolution manners. The singular values of each convolution scheme is normalized by its largest value.}}
   \label{fig:svd}
\end{figure}

\subsection{More Analysis}
\label{sec:discussion}

%\st{Some low-dimensional feature extraction methods \cite{Dong2019Deep, Li2020Mixed, Li2021Exploring} have also been utilized for HS image processing tasks,
%\st{Although comparable results can be achieved by these methods},
%most of them empirically applied a 2-D spatial convolution and 1-D spectral convolution to perform HS feature extraction, which is widely employed in video processing tasks \cite{Qiu2017Learning,xie2018rethinking}.}

%\textcolor{magenta}{
We also analyze the existing low-dimensional convolution-based feature extraction manners by using our formulation from the perspective of feature diversity %the above-mentioned feature diversity-based analysis
to deeply understand them.
%theoretical limitations of those methods.
%}
Particularly, we take ``Sequential 1-D and 2-D convolution" in Fig. \ref{fig:low-rank-filters} (b) and ``1-D + 2-D convolution" in Fig. \ref{fig:low-rank-filters} (d)  as examples, and refer the readers to Table \ref{tab:differentfilters} for the  quantitative results of all the other variants illustrated in Fig. \ref{fig:low-rank-filters}.
%\st{Besides, they fail to theoretically investigate the interpretability of designing the feature extraction module.
%The feature extraction operation of these schemes are shown in Figs. \ref{fig:low-rank-filters} (b).
%Although such approaches could achieve better performance than the methods using the original 3-D filter as feature extraction module, they fail to theoretically investigate the interpretability of designing the feature extraction.
%Besides, such approaches fail to theoretically design the feature extraction module and optimize the architectures, which may be not optimal, thus compromising performance.
%By contrast, we design our low-dimensional HS feature extraction module by quantitatively analyzing the feature diversity, resulting in efficient and effective feature learning, and parameter reduction.}

%In particular, such a feature extraction module with 2-D and 1-D convolutions can be also explored by our quantitative analysis of feature diversity.
%In our settings, the feature extraction operation using 2-D and 1-D convolutions
Specifically, %for ``1-D + 2D convolution",
we can equivalently write their form of matrix multiplication as
%can be formulated as:\vspace{-0.2cm}
\begin{equation}
\begin{split}
\label{equ:1d2d-conv2}
\mathbf{F}_{e1} = \mathbf{A}_{e2} \cdot \mathbf{I}_{2D}', ~~
\mathbf{I}_{2D} = \mathbf{A}_{e1} \cdot \mathbf{I},
\end{split}
\end{equation}
and
\begin{equation}
\begin{split}
\label{equ:1d2d-conv1}
\mathbf{F}_{e2} = \mathbf{A}_{e} \cdot \mathbf{I}, ~~~{\rm with}~~
\mathbf{A}_{e} = [\mathbf{A}_{e3};\mathbf{A}_{e4}],
\end{split}
\end{equation}
where
$\mathbf{F}_{e1}\in \mathbb{R}^{M \times B'H'W'}$ and $\mathbf{F}_{e2}\in \mathbb{R}^{2M \times B'H'W'}$ encode the output feature maps from ``Sequential 1-D and 2-D convolution" and ``1-D + 2-D convolution", respectively;
$\mathbf{A}_{e1}\in \mathbb{R}^{M \times k^3C}, \mathbf{A}_{e3}\in \mathbb{R}^{M \times k^3C}$ and $\mathbf{A}_{e2}\in \mathbb{R}^{M \times k^3M}, \mathbf{A}_{e4}\in \mathbb{R}^{M \times k^3C}$ are the corresponding matrices of 2-D and 1-D kernels, respectively;
%$\mathbf{C}_{e}\in \mathbb{R}^{M \times 2M}$ is the matrix of concatenation layer with $1\times 1\times 1$ convolutional filter.
$\mathbf{I}_{2D} \in \mathbb{R}^{M \times B'H'W'}$ is the matrix form of intermediate feature maps; $\mathbf{I}_{2D}' \in \mathbb{R}^{k^3M \times B'H'W'}$ is generated by sliding 1-D kernels across the intermediate features;
$\mathbf{A}_{e}\in \mathbb{R}^{2M \times k^3C}$ denotes the  matrix of combined 1-D and 2-D kernels, which is illustrated in Fig.~\ref{fig:filter-rank} (b).
%The ``1-D + 2-D" and ``Sequential 1-D and 2-D" convolutional kernels are
%illustrated in Fig.~\ref{fig:filter-rank} (b) and (c), respectively.
%Similarly, a $1\times 1\times 1$ convolutional layer is applied to compress the output feature.
%The feature extraction process of this scheme is shown in Fig. \ref{fig:low-rank-filters} (b).

According to Eqs. (\ref{equ:1d2d-conv2}) and (\ref{equ:1d2d-conv1}),
%the rank of matrix $\mathbf{F}_{e}$ can be obtained as
we have $\mathsf{Rank}(\mathbf{F}_{e1})\leq M$ and $\mathsf{Rank}(\mathbf{F}_{e2})\leq 2M$. That is, such feature extraction manners expand the rank upper bound
%of feature diversity
from $M$ to $2M$ at most, which is still limited. %compared with the original 3-D kernel.
%Thus, the rank of matrix $\mathbf{A}_{e}$ can be obtained as $\mathsf{Rank}(\mathbf{A}_{e})\leq 2M$, which shows great potential in expanding the filter diversity compared with the original 3-D filter.
%This could be the reason why this feature extraction module outperforms the original 3-D feature extraction module.
%However, compared with our Re-ConvSet, the feature diversity is still limited. %See Sec.~\ref{subsubsec:abexp-filterrank} for
%\st{more analysis and}
%the quantitative comparisons toward different feature extraction manners.
\begin{table*}[t]
\caption{Quantitative comparisons of different methods under several noise levels over the ICVL dataset. The best and second best results are highlighted in bold and underline, respectively. ``$\uparrow$" (resp. ``$\downarrow$") means the larger (resp. smaller), the better.}
\centering
\begin{spacing}{1.05}
\label{tab:icvlgaussianresults}
\resizebox{0.95\textwidth}{!}{
\begin{tabu}{c|c|c|c|c|c|c|c|c|c|c}
\tabucline[0.5pt]{*}
\hline
\multirow{3}{*}{$\sigma$}        &\multirow{3}{*}{Metrics}  &\multicolumn{9}{c}{Methods}                         \\ \cline{3-11}
~ &~ &Noisy &BM4D &TDL &ITSReg &LLRT &GRN &QRNN3D &T3SC &Ours \\ ~ &~ &~ &\cite{Maggioni2013Nonlocal} &\cite{Peng2014Decomposable} & \cite{Xie2016Multispectral} &\cite{Chang2017Hyper} &\cite{Cao2021Deep} &\cite{Wei20213D}  &\cite{Bodrito2021a}  &~       \\  \hline\hline
%\multirow{3}{*}{-}        &\#Params (M)        &-     &-     &-    &-    &-     &-    &-   &-      \\
%                          &\#FLOPs  (T)        &-     &-     &-    &-    &-     &-    &-   &-      \\
%                          &\#Running time (ms) &-     &-     &-    &-    &-     &-    &-   &-      \\ \hline
\multirow{3}{*}{30}        &MPSNR$\uparrow$  &18.59   &38.29  &40.87   &41.53    &41.83    &41.52     &42.28     &\underline{43.19}    &\textbf{43.68}   \\ %\cline{2-11}
                           &MSSIM$\uparrow$  &0.1034  &0.9342 &0.9557  &0.9571   &0.9653   &0.9698    &0.9701    &\underline{0.9718}   &\textbf{0.9747}  \\
                           &SAM$\downarrow$  &0.7269  &0.1177 &0.0634  &0.0929   &\underline{0.0541}  &0.0690    &0.0617   &0.0616     &\textbf{0.0448}  \\ \hline
\multirow{3}{*}{50}        &MPSNR$\uparrow$  &14.15   &35.54  &38.50   &39.19    &38.84    &39.86     &40.22     &\underline{40.81}    &\textbf{41.39}   \\ %\cline{2-11}
                           &MSSIM$\uparrow$  &0.0429  &0.8929 &0.9323  &0.9350   &0.9422   &0.9535    &0.9544    &\underline{0.9567}   &\textbf{0.9607}  \\          &SAM$\downarrow$  &0.9096  &0.1535 &0.0841  &0.1010   &0.0734   &0.0810    &0.0733    &\underline{0.0720}   &\textbf{0.0534}  \\ \hline
\multirow{3}{*}{70}        &MPSNR$\uparrow$  &11.23   &33.71  &36.91   &37.48    &37.22    &38.27     &38.29     &\underline{39.27}    &\textbf{39.85}   \\ %\cline{2-11}
                           &MSSIM$\uparrow$  &0.0228  &0.8545 &0.9104  &0.9192   &0.9264   &0.9333    &0.9326    &\underline{0.9431}   &\textbf{0.9478}  \\
                           &SAM$\downarrow$  &1.0273  &0.1815 &0.1002  &0.1144   &0.0853   &0.0933    &0.0943    &\underline{0.0810}   &\textbf{0.0606}  \\ \hline
\multirow{3}{*}{Blind}     &MPSNR$\uparrow$  &14.83   &35.94  &38.86   &39.52    &39.21    &40.06     &40.48     &\underline{40.98}    &\textbf{41.72}   \\ %\cline{2-11}
                           &MSSIM$\uparrow$  &0.0534  &0.8979 &0.9353  &0.9389   &0.9452   &0.9551    &0.9559    &\underline{0.9579}   &\textbf{0.9626}  \\
                           &SAM$\downarrow$  &0.8800  &0.1484 &0.0828  &0.1037   &\underline{0.0717}  &0.0801    &0.0737     &0.0723   &\textbf{0.0527}  \\ \hline

\tabucline[0.5pt]{*}
\end{tabu}}
\end{spacing}
\end{table*}

\begin{table*}[t]
\caption{Quantitative comparisons of different methods under five complex noise cases over the \textbf{ICVL} dataset. %The best and second best results are highlighted in bold and underline, respectively.
The best and second best results are highlighted in bold and underline, respectively. ``$\uparrow$" (resp. ``$\downarrow$") means the larger (resp. smaller), the better.
}
\centering
\begin{spacing}{1.05}
\label{tab:icvlcomplexresults}
\resizebox{0.95\textwidth}{!}{
\begin{tabu}{c|c|c|c|c|c|c|c|c|c|c}
\tabucline[0.5pt]{*}
\hline
\multirow{3}{*}{Case}        &\multirow{3}{*}{Metrics}  &\multicolumn{9}{c}{Methods}                           \\ \cline{3-11}
~ &~ &Noisy &LRMR &LRTV &NMoG &TDTV &GRN  &QRNN3D &T3SC &Ours  \\
~ &~ &~ &\cite{Zhang2014Hyperspectral} &\cite{He2016Total} &\cite{Chen2018Denoising} &\cite{Wang2018Hyperspectral} &\cite{Cao2021Deep} &\cite{Wei20213D}  &\cite{Bodrito2021a} &~ \\  \hline\hline
%\multirow{3}{*}{-}         &\#Params (M)        &-     &-     &-    &-    &-     &-    &-   &-      \\
%                           &\#FLOPs  (T)        &-     &-     &-    &-    &-     &-    &-   &-      \\
%                           &\#Running time (ms) &-     &-     &-    &-    &-     &-    &-   &-      \\ \hline
\multirow{3}{*}{1}         &MPSNR$\uparrow$  &17.80   &28.64   &33.96   &34.96   &37.95   &39.97   &42.79   &\underline{43.51}   &\textbf{44.25}  \\ %\cline{2-11}
                           &MSSIM$\uparrow$  &0.1516  &0.5153  &0.8987  &0.8279  &0.9377  &0.9587  &0.9752  &\underline{0.9776}  &\textbf{0.9799} \\ %\cline{2-11}
                           &SAM$\downarrow$  &0.7911  &0.3235  &0.0647  &0.1260  &0.0671  &0.0685  &\underline{0.0430}   &0.0441 &\textbf{0.0330} \\ \hline
\multirow{3}{*}{2}         &MPSNR$\uparrow$  &17.77   &28.52   &34.05   &34.60   &37.65   &39.90   &42.64   &\underline{43.20}   &\textbf{44.28}  \\ %\cline{2-11}
                           &MSSIM$\uparrow$  &0.1545  &0.5155  &0.8996  &0.8184  &0.9348  &0.9598  &0.9750  &\underline{0.9770}  &\textbf{0.9804} \\ %\cline{2-11}
                           &SAM$\downarrow$  &0.7895  &0.3250  &0.0662  &0.1793  &0.0731  &0.0672  &\underline{0.0437}  &0.0488  &\textbf{0.0323} \\ \hline
\multirow{3}{*}{3}         &MPSNR$\uparrow$  &17.36   &27.78   &32.72   &33.60   &35.67   &38.74   &\underline{42.31}   &41.42   &\textbf{44.27}  \\ %\cline{2-11}
                           &MSSIM$\uparrow$  &0.1473  &0.5075  &0.8908  &0.8212  &0.9181  &0.9548  &\underline{0.9735}  &0.9724  &\textbf{0.9801} \\ %\cline{2-11}
                           &SAM$\downarrow$  &0.8109  &0.3398  &0.1021  &0.1885  &0.0937  &0.0702  &\underline{0.0455}  &0.0639  &\textbf{0.0332} \\ \hline
\multirow{3}{*}{4}         &MPSNR$\uparrow$  &14.86   &24.19   &32.41   &29.09   &36.60   &37.63   &\underline{40.49}   &37.93   &\textbf{42.31}  \\ %\cline{2-11}
                           &MSSIM$\uparrow$  &0.1118  &0.3805  &0.8722  &0.6751  &0.9265  &0.9410  &\underline{0.9533}  &0.9353  &\textbf{0.9641} \\ %\cline{2-11}
                           &SAM$\downarrow$  &0.8480  &0.4681  &0.1983  &0.4510  &0.0874  &0.0952  &\underline{0.0762}  &0.1669  &\textbf{0.0650} \\ \hline
\multirow{3}{*}{5}         &MPSNR$\uparrow$  &14.07   &23.79   &31.39   &28.45   &34.51   &38.01   &\underline{39.42}   &35.84   &\textbf{41.77}  \\ %\cline{2-11}
                           &MSSIM$\uparrow$  &0.0936  &0.3817  &0.8649  &0.6746  &0.9076  &\underline{0.9473}  &0.9448  &0.9248  &\textbf{0.9615} \\ %\cline{2-11}
                           &SAM$\downarrow$  &0.8587  &0.4668  &0.2135  &0.4568  &0.1063  &0.0904  &\underline{0.0809}  &0.1804  &\textbf{0.0673} \\ \hline
\tabucline[0.5pt]{*}
\end{tabu}}
\end{spacing}
\end{table*}

\emph{Remarks.} For %those
the feature extraction manner involving parallel branches, e.g., Figs. \ref{fig:low-rank-filters} (d) and (e), we utilize a $1\times 1\times 1$ convolutional layer to compress the multiple output feature volumes before feeding it into the subsequent layer in order to avoid channel explosion. %, leading to that
Thus, the feature volumes extracted by different convolutional manners are finally with the equal size,
%Although the feature maps extracted by different convolutional manners are finally compressed into the same volume,
i.e., the rank upper bounds of finally-output feature matrices are equal.
However, %it is worth noting that
our Re-ConvSet can boost the rank upper bound of feature matrix $\mathbf{F}$ from $M$ to $3M$ during the feature extraction process, thus potentially promoting more diverse features. Here,
we compared the singular value distributions of the feature maps extracted by various convolution manners in Fig. \ref{fig:svd}, where it can be clearly seen that the singular values of the feature matrix by our Re-ConvSet decrease more slowly than those of other schemes,
indicating that our Re-ConvSet can balance the singular values to avoid only a few large ones dominating the feature space (i.e., the degree of freedom of entries of the feature matrix that is approximately low-rank is limited), thus promoting feature diversity.

\section{Experiments}
\label{sec:experiments}

\subsection{Experiment Settings}

\subsubsection{Datasets} We employed four commonly-used HS image benchmark datasets for evaluation, including two natural HS image datasets, i.e., ICVL\footnote{http://icvl.cs.bgu.ac.il/hyperspectral/} \cite{Arad2016Sparse} and  CAVE\footnote{http://www.cs.columbia.edu/CAVE/databases/} \cite{Yasuma2010CAVE}, and two remote sensing HS images, i.e.,
%Pavia Center\footnote{http://www.ehu.eus/ccwintco/index.php/Hyperspectral\_Remote\_Sensing\_\\Scenes/\label{web}},
Pavia University\footnote{http://www.ehu.eus/ccwintco/index.php/Hyperspectral\_Remote\_\\Sensing\_Scenes/\label{web}}
%Indian Pines\footnote{https://engineering.purdue.edu/ biehl/MultiSpec/hyperspectral.html},
and Urban\footnote{https://rslab.ut.ac.ir/data}, whose details are listed as follows:
%\textcolor{red}{[This part can be simplified if the space is not enough]}

\begin{itemize}
  \item The ICVL dataset consists of 201 HS images of spatial dimensions $1392\times1300$ and spectral dimension 31 covering the wavelength in the range of 400 to 700 nm, acquired by a Specim PS Kappa DX4 HS camera. We utilized 100 HS images as the training set, and the rest as the testing set.
  \item The CAVE dataset contains 32 HS images of spatial dimensions $512\times512$ and spectral dimension 31 covering the wavelength in the range of 400 to 700 nm, collected by a generalized assorted pixel camera. Note that we randomly selected 10 HS images from this dataset \textit{only for testing}.
  \item Pavia University contains %one HS image of spatial dimensions
  $610\times 610$ pixels and $103$ spectral bands gathered by the ROSIS sensor. This image is only used for testing.
  \item Urban contains $307\times307$ pixels and 210 spectral bands collected by the HYDICE hyperspectral system. This image is corrupted by \textit{real unknown noise} and widely used for real HS image denoising testing.
  %Both of these two HS datasets are widely used for real HS image denoising testing \cite{}.
\end{itemize}
%\subsubsection{Noise settings.}
%\st{Real HS images are usually corrupted, due to suffering from several kinds of noise, such as the most common Gaussian noise, stripes, deadlines, impulse noise, or a mixed one of them. Thus,}
Following previous works \cite{Wei20213D,Cao2021Deep}, we considered two kinds of noise settings, i.e., the Gaussian noise and the complex noise, which were applied to ICVL and CAVE datasets and Pavia University to simulate noisy HS images. % conducted two kinds of denoising experiments, %to evaluate the performance of our Re-ConvSet, including
Specifically, for the Gaussian noise, %denoising experiments were conducted under different
we set various noise levels, i.e., $\sigma=30$, $50$, $70$, and ``Blind (the value of $\sigma$ is in the range of 30 to 70 but unknown)".
%"Blind" means suggests each sample is corrupted by Gaussian noise with unknown $\sigma$ (range from 30 to 70).
%For the complex noise,
We generated five types of complex noises to imitate the real-world noise cases, including Non-i.i.d. Gaussian Noise, Gaussian and Stripe Noise, Gaussian and Deadline Noise, Gaussian and Impulse Noise, and Mixture Noise, referred as ``Case 1" to ``Case 5". We refer the readers to \cite{Wei20213D,Cao2021Deep} for more details about the noise settings

\begin{figure*}[!t]
\centering
\includegraphics[width=0.95\linewidth]{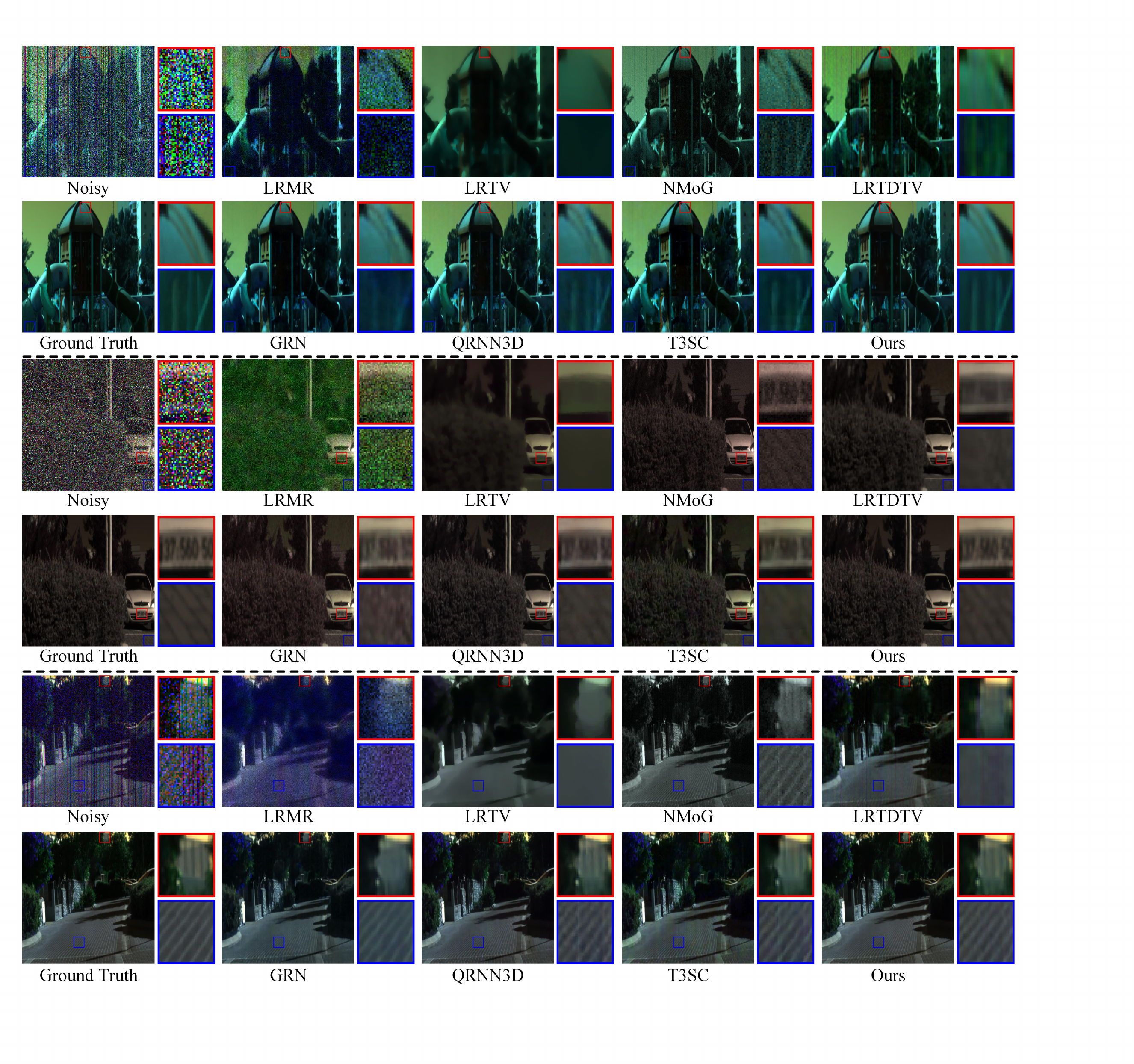}
\caption{Visual comparison of different methods on \textit{ICVL} with Gaussian and deadline, Gaussian and impulse, and mixture noise, from top to bottom. Here, we selected the $5^{th}$, $16^{th}$, and $31^{st}$ (resp. $24^{th}$, $28^{th}$, and $30^{th}$, and $10^{th}$, $17^{th}$, and $30^{th}$) bands to form a pseudo RGB image to enable the visualization under Gaussian and deadline noise (resp. Gaussian and impulse, and mixture noise). }
\label{fig:icvl-complex}
\end{figure*}

\begin{figure*}[!t]
\centering
\includegraphics[width=1\linewidth]{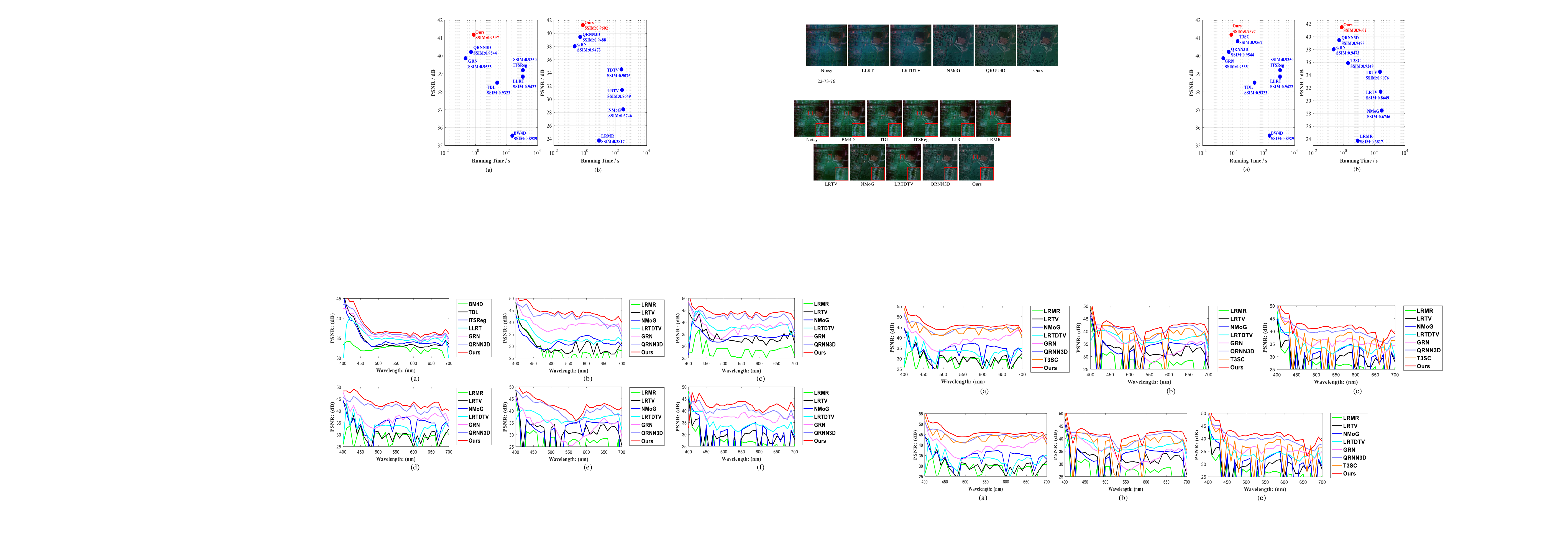}
\caption{PSNR values across the spectrum corresponding to complex noise removal results in Fig \ref{fig:icvl-complex}, respectively. (a) Gaussian and Deadline (Case 3), (b) Gaussian and Impulse (Case 4), and (c) Mixture noise (Case 5).
}
\label{fig:psnr-across-bands}
\end{figure*}

\begin{table}[t]
\caption{Comparisons of \#Param and \#FLOPs of deep learning-based methods over the ICVL dataset. %For \#Param and \#FLOPs, the smaller, the more compact and efficient.
Since T3SC \cite{Bodrito2021a} was built based on sparse coding and deep learning, we could not calculate the \#FLOPs like other pure deep learning-based methods.
}
\centering
\begin{spacing}{1.05}
\label{tab:time-flops}
\resizebox{0.45\textwidth}{!}{
\begin{tabu}{c|c|c|c|c}
\tabucline[0.5pt]{*}
\hline
\multirow{2}{*}{Metrics}  &\multicolumn{4}{c}{Methods}                            \\ \cline{2-5}
~ &~GRN \cite{Cao2021Deep}~  &~QRNN3D \cite{Wei20213D}~ &~T3SC \cite{Bodrito2021a}~   &~~Ours~~ \\  \hline\hline
\#Param (M)      &1.07  &0.86   &0.83   &0.66  \\ \cline{1-5}
\#FLOPs (T)      &0.22  &1.26   &N/A    &0.95  \\  \hline
\tabucline[0.5pt]{*}
\end{tabu}}
\end{spacing}
\end{table}

\begin{table*}[t]
\caption{Quantitative comparisons of different methods on the \textbf{CAVE} dataset under five complex noise scenarios. The best and second best results are highlighted in bold and underline, respectively. ``$\uparrow$" (resp. ``$\downarrow$") means the larger (resp. smaller), the better.
}
\centering
\begin{spacing}{1.05}
\label{tab:caveresults}
\resizebox{0.95\textwidth}{!}{
\begin{tabu}{c|c|c|c|c|c|c|c|c|c|c}
\tabucline[0.5pt]{*}
\hline
\multirow{3}{*}{Case}        &\multirow{3}{*}{Metrics}  &\multicolumn{9}{c}{Methods}                           \\ \cline{3-11}
~ &~ &Noisy &LRMR &LRTV &NMoG &TDTV &T3SC &GRN  &QRNN3D  &Ours  \\
~ &~ &~ &\cite{Zhang2014Hyperspectral} &\cite{He2016Total} &\cite{Chen2018Denoising} &\cite{Wang2018Hyperspectral} &\cite{Bodrito2021a} &\cite{Cao2021Deep} &\cite{Wei20213D} &~ \\  \hline\hline
%\multirow{3}{*}{-}   &\#Params (M)        &-     &-     &-    &-    &-     &-    &-   &-      \\
%                     &\#FLOPs  (T)        &-     &-     &-    &-    &-     &-    &-   &-      \\
%                     &\#Running time (ms) &-     &-     &-    &-    &-     &-    &-   &-      \\ \hline
\multirow{3}{*}{1}   &MPSNR$\uparrow$  &17.48   &29.25   &31.84   &32.09    &34.80   &36.89   &37.23     &\underline{39.25}     &\textbf{40.68}  \\ %\cline{2-11}
                     &MSSIM$\uparrow$  &0.1473  &0.5641  &0.9003  &0.7481   &0.9180  &0.9226   &0.9395   &\underline{0.9558}    &\textbf{0.9730}   \\ %\cline{2-11}
                     &SAM$\downarrow$  &1.1911  &0.8073  &0.2821  &0.5559   &0.3270  &0.2530   &\underline{0.2225}     &0.2672  &\textbf{0.2092}   \\ \hline
\multirow{3}{*}{2}   &MPSNR$\uparrow$  &17.55   &29.28   &31.88   &31.95    &34.68   &36.72    &37.18    &\underline{39.31}     &\textbf{40.75}  \\ %\cline{2-11}
                     &MSSIM$\uparrow$  &0.1524  &0.5694  &0.8982  &0.5736   &0.9144  &0.9183   &0.9401   &\underline{0.9546}    &\textbf{0.9729}   \\ %\cline{2-11}
                     &SAM$\downarrow$  &1.1861  &0.7920  &0.3176  &0.5736   &0.3339  &0.2589   &\underline{0.2225}  &0.2829     &\textbf{0.2095}   \\ \hline
\multirow{3}{*}{3}   &MPSNR$\uparrow$  &17.28   &28.52   &31.00   &31.56    &34.30   &36.45    &36.94    &\underline{38.96}     &\textbf{40.64}  \\ %\cline{2-11}
                     &MSSIM$\uparrow$  &0.1476  &0.5627  &0.8829  &0.7574   &0.9099  &0.9149   &0.9262   &\underline{0.9500}    &\textbf{0.9723}   \\ %\cline{2-11}
                     &SAM$\downarrow$  &1.1996  &0.8026  &0.3399  &0.5943   &0.3430  &0.2614   &\underline{0.2356}    &0.2797   &\textbf{0.2214}   \\ \hline
\multirow{3}{*}{4}   &MPSNR$\uparrow$  &14.39   &24.48   &30.19   &26.57    &34.51   &33.20    &35.73    &\underline{36.65}     &\textbf{38.10}  \\ %\cline{2-11}
                     &MSSIM$\uparrow$  &0.1067  &0.4837  &0.8377  &0.6007   &\underline{0.9076} &0.8331  &0.8811     &0.8814    &\textbf{0.9104}   \\ %\cline{2-11}
                     &SAM$\downarrow$  &1.1224  &0.8153  &0.5740  &0.8136   &0.3431  &0.4729    &0.3265  &\underline{0.3193}    &\textbf{0.3071}   \\ \hline
\multirow{3}{*}{5}   &MPSNR$\uparrow$  &13.77   &23.83   &29.26   &25.23    &33.56   &31.74     &34.57   &\underline{35.31}     &\textbf{37.19}    \\ %\cline{2-11}
                     &MSSIM$\uparrow$  &0.0974  &0.4749  &0.8091  &0.5718   &\textbf{0.9013}    &0.8019  &0.8413     &0.8505    &\underline{0.8831}   \\ %\cline{2-11}
                     &SAM$\downarrow$  &1.1438  &0.8506  &0.6138  &0.8168   &\underline{0.3418} &0.5300  &\textbf{0.3324}  &0.3900    &0.3551    \\ \hline
\tabucline[0.5pt]{*}
\end{tabu}}
\end{spacing}
%\vspace{-0.1cm}
\end{table*}

\subsubsection{Implementation details} We implemented all the experiments %are implemented
with PyTorch on a machine with NVIDIA GeForce RTX 3080 GPU, Intel(R) Core(TM) i7-10700 CPU of 2.90GHz and 64-GB RAM.
%\textcolor{red}{[list the GPU model and RAM!!]}.
%We embedded the proposed Re-ConvSet into the widely-used U-Net, and utilized the $L_1$ loss as the objective function.
We employed the ADAM optimizer \cite{kingma2014adam} with the exponential decay rates $\beta_1=0.9$ and $\beta_2=0.999$. The total training process was 25 epochs for both two kinds of noise experiments. We initialized the learning rate as $5\times10^{-4}$, which was halved every 5 epochs. We set the batch size to 4 in all experiments.
%\textcolor{red}{[I think the following sentence can be removed first]} Finally, two models were trained to handle all the kinds of noise for the Gaussian and complex noise cases, respectively.\vspace{-0.3cm}

\subsubsection{Compared methods} We compared the proposed denoising method with three state-of-the-art deep learning-based methods, i.e., QRNN3D \cite{Wei20213D}, GRN \cite{Cao2021Deep}, and T3SC \cite{Bodrito2021a}, and eight representative traditional methods, including BW4D \cite{Maggioni2013Nonlocal}, TDL \cite{Peng2014Decomposable}, ITSReg \cite{Xie2016Multispectral}, LLRT \cite{Chang2017Hyper} for the Gaussian noise scenario, and LRMR \cite{Zhang2014Hyperspectral}, LRTV \cite{He2016Total}, NMoG \cite{Chen2018Denoising}, TDTV \cite{Wang2018Hyperspectral} for the complex noise scenario.
For a fair comparison, we  retrained %\st{re-implemented or fine-tuned}
all the compared deep learning-based methods with the same training data as ours and the codes released by the authors with suggested settings. Note that for GRN, QRNN3D, and our method, a single model was trained to handle various Gaussian noise levels (resp. complex noise cases), while for T3SC, a model was trained for each Gaussian noise level (resp. complex noise case). Thus, the setting favors T3SC.

\begin{table*}[t]
\caption{Quantitative comparisons of different methods on \textit{Pavia University} with the mixture noise. The best and second best results are highlighted in bold and underline, respectively. ``$\uparrow$" (resp. ``$\downarrow$") means the larger (resp. smaller), the better.
}
\centering
\begin{spacing}{1.0}
\label{tab:paviaresults}
\resizebox{0.9\textwidth}{!}{
\begin{tabu}{c|c|c|c|c|c|c|c}
\tabucline[1pt]{*}
\hline
\multirow{2}{*}{Metrics}  &\multicolumn{7}{c}{Methods}                           \\ \cline{2-8}
~ &Noisy &LRMR \cite{Zhang2014Hyperspectral} &LRTV \cite{He2016Total} &NMoG \cite{Chen2018Denoising} &TDTV \cite{Wang2018Hyperspectral}  &QRNN3D \cite{Wei20213D}  &Ours        \\  \hline\hline
%\multirow{3}{*}{-}         &\#Params (M)        &-     &-     &-    &-    &-     &-    &-   &-      \\
%                           &\#FLOPs  (T)        &-     &-     &-    &-    &-     &-    &-   &-      \\
%                           &\#Running time (ms) &-     &-     &-    &-    &-     &-    &-   &-      \\ \hline
MPSNR$\uparrow$     &13.63  &26.47 &26.76  &29.33  &30.56  &\underline{33.27}  &\textbf{33.87}   \\ %\cline{1-8}
MSSIM$\uparrow$     &0.1544     &0.6496     &0.6812     &0.7847     &0.8170         &\underline{0.9062}    &\textbf{0.9162}    \\ %\cline{1-8}
SAM$\downarrow$     &0.8995     &0.4097     &0.3087     &0.4180     &0.2646         &\underline{0.1043}    &\textbf{0.0865}    \\ \hline
\tabucline[1pt]{*}
\end{tabu}}
\end{spacing}
\end{table*}

\begin{figure*}[!t]
\centering
\includegraphics[width=1\linewidth]{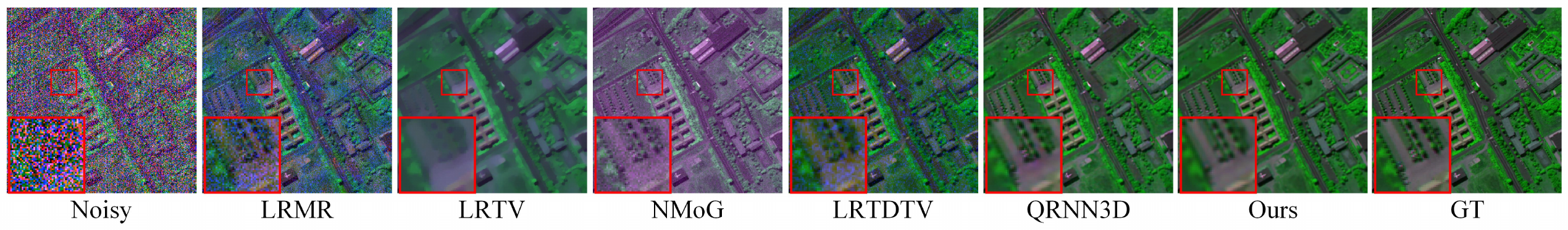}
\caption{Visual comparison of different methods on \textit{Pavia University} with the mixture noise. Here, we selected the $43^{rd}$, $66^{th}$, and $103^{rd}$ bands to form a pseudo RGB image to enable the visualization.
%For ease of comparison, we visualized the denoised results with three spectral bands.
}\vspace{-0.3cm}
\label{fig:paviau}
\end{figure*}

\subsubsection{Evaluation metrics} Following previous works \cite{Wei20213D,Cao2021Deep}, we adopted three commonly used quantitative metrics to evaluate the quality of the denoised HS images, %\st{$\widehat{\mathbf{Y}}$ quantitatively},
i.e., Mean Peak Signal-to-Noise Ratio (MPSNR), Mean Structural Similarity Index (MSSIM) \cite{Zhou2002SSIM}, and Spectral Angle Mapper (SAM) \cite{Yuhas1992Discrimination}. For MPSNR and MSSIM, the larger, the better. For SAM, the smaller, the better.

\subsection{Evaluation on Natural HS Images} %the ICVL Dataset
Tables \ref{tab:icvlgaussianresults} and \ref{tab:icvlcomplexresults} show the quantitative results of different methods applied to handle HS images with the Gaussian noise and complex noise, respectively, where it can be observed that
\begin{itemize}
   \item our method consistently achieves the best performance in terms of all the three metrics under all noise scenarios.
  % validating the effectiveness and advantage of our Re-ConvSet.
   Particularly, our method improves the MPSNR of the second best methods by $0.49$ dB, $0.58$ dB, $0.58$ dB, and $0.74$ dB under four Gaussian noise levels, and $0.74$ dB, $1.08$ dB, $1.96$ dB, $1.82$ dB, and $2.35$ dB under five types of complex noises, respectively;

  \item the impressive performance of our method under %performs very well on the
  both the blind Gaussian noise  and all the complex noise scenarios demonstrates
  that it has better resilience on severely corrupted HS images; and

  %\item
      % in gaussian noise case, traditional denoising methods, such as TDL \cite{Peng2014Decomposable}, ITSReg \cite{Xie2016Multispectral}, and LLRT \cite{Chang2017Hyper} can achieve comparable performance compared with the state-of-the-art deep learning-based methods, i.e., GRN \cite{Cao2021Deep} and QRNN3D \cite{Wei20213D}. Especially, LLRT \cite{Chang2017Hyper} even obtain better denoising results than these two methods in terms of SAM. By contrast, our REF-Net significantly outperforms all the traditional methods;

  \item existing deep learning-based methods, i.e., GRN \cite{Cao2021Deep}, QRNN3D \cite{Wei20213D}, and T3SC \cite{Bodrito2021a}, surpass the traditional denoising methods to a large extent under the complex noise scenario. Moreover, our method further boosts their performance. %, demonstrating the superiority of our Re-ConvSet.
\end{itemize}

%Besides, %Fig. \ref{fig:icvl-guass} and
Fig. \ref{fig:icvl-complex} shows the visual comparisons of denoising results by different methods, demonstrating the advantage of our method again, where we can observe that the denoised images by our method is cleaner and retain the
%produces better visual results with noise removing and while preserving
original high-frequency details better.
%of the original structure-preserving simultaneously. %which further demonstrates its advantage.
In addition, Fig. \ref{fig:psnr-across-bands} illustrates the PSNR value of each band of the denoised HS images %Gaussian and complex noise
%the denoising results
shown in Fig. \ref{fig:icvl-complex}, where it can be seen that our method achieves the highest PSNR values on almost all spectral bands.
%can obviously achieve higher PSNR values across the spectrum than all the compared methods. %demonstrating the advantage of our proposed Re-ConvSet.

%\subsubsection{Computational efficiency.}
Besides, we also compared %the computational efficiency
the number of network parameters (\#Param) and floating point of operations (\#FLOPs) of deep learning-based methods
%measured with the number of neural network parameters (\#Param), floating point of operations (\#FLOPs) %, \st{and running time of per image}
in Table \ref{tab:time-flops}, where it can be observed that our method consumes fewer network parameters and has comparable \#FLOPs,
%and running time
compared with state-of-the-art methods, demonstrating that the excellent performance of our method does not come at the cost of a larger capacity and  higher computational complexity but is credited to elegant feature extraction technique.

Finally, Table \ref{tab:caveresults} lists  the quantitative results of different methods on the CAVE dataset under the five complex noise scenarios, where for the deep learning-based methods, we directly applied the models trained on the ICVL dataset. From Table \ref{tab:caveresults}, it can be seen that our method achieves superior denoising  performance to other state-of-the-art methods, demonstrating its strong generalization ability. % of our Re-ConvSet.

\subsection{Evaluation on Remote Sensing HS Images}

\begin{figure*}[!t]
\centering
\includegraphics[width=0.95\linewidth]{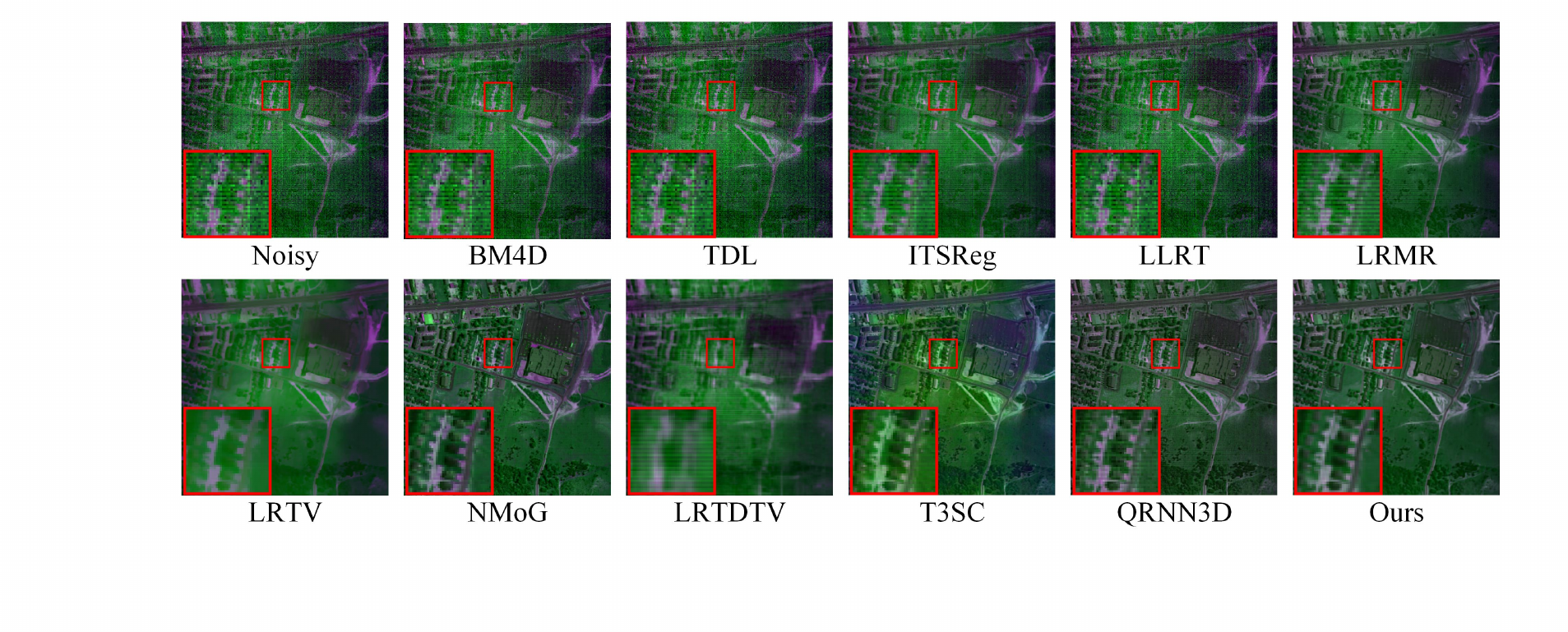}
\caption{Visual comparison of different methods on \textit{Urban} with real unknown noise. Here we selected the $38^{th}$, $112^{nd}$, and $128^{th}$ bands to form a pseudo RGB image to enable the visualization.
}
\label{fig:urban}
\end{figure*}

\begin{table*}[t]
\caption{Quantitative comparisons of different combinations of low-dimensional convolution over the ICVL dataset with the Gaussian noise ($\sigma=70$). The best result are highlighted in bold. %\textcolor{red}{[Update the results of 2-D + 3-D conv.!!]}
}
\centering
\label{tab:differentfilters}
\resizebox{\textwidth}{!}{
\begin{tabu}{c|c|c|c|c|c|c}  %
\tabucline[1pt]{*}
Methods  &\#Params (M) &\#FLOPs (T)    &Rank upper bound     &MPSNR$\uparrow$   &MSSIM$\uparrow$   &SAM$\downarrow$    \\  \hline\hline  %\hline
3-D conv. (Fig. \ref{fig:low-rank-filters} (a))% (baseline)
&1.201    &1.562    &$M$     &39.34       &0.9430   &0.0693      \\ %\hline
Seq. 1-D conv. (Fig. \ref{fig:low-rank-filters} (c))%(w/o rank-enhancing)
&0.614    &0.765    &$M$     &39.42          &0.9430   &0.0658      \\
Seq. 1-D and 2-D conv. (Fig. \ref{fig:low-rank-filters} (b))%(w/o rank-enhancing)
&0.717    &0.889    &$M$     &39.58           &0.9445   &0.0652      \\
%2-D + 3-D conv. (Fig. \ref{fig:low-rank-filters} (e)) %(w rank-enhancing)
%&1.533    &2.355    &$2M$    &42.87          &0.9718   &0.0484      \\
1-D + 2-D conv. (Fig. \ref{fig:low-rank-filters} (d)) %(w rank-enhancing)
&0.740    &1.082    &$2M$     &39.71           &0.9459   &0.0610      \\
%1-D low-dim. filters (w rank-enhancing)
Re-ConvSet &0.658    &0.953    &$3M$     &\textbf{39.85}  &\textbf{0.9478}   &\textbf{0.0606}      \\ \hline
\tabucline[1pt]{*}
\end{tabu}}
%\vspace{-0.3cm}
\end{table*}

To demonstrate the generalization ability of the proposed method, we also conducted experiments on remote sensing HS images, whose spectral characteristics are significantly different from those of the previous natural HS images. Note that for all the deep learning-based methods, we directly applied the models trained on the ICVL dataset under this scenario. %\vspace{-0.3cm}
\subsubsection{Synthetic noisy data}
Table \ref{tab:paviaresults} shows the results of different methods on Pavia University with the synthetic mixture noise, where %for the deep learning-based methods,
%\st{The results of deep learning-based methods are also obtained by directly utilizing}
%we directly applied the models trained on the ICVL dataset.
as GRN \cite{Cao2021Deep} and T3SC \cite{Bodrito2021a} cannot handle the  HS images with a different number of spectral bands from the training data, we did not report their results under this scenario.
%in Table \ref{tab:paviaresults} and Fig. XXXXX.
From Table \ref{tab:paviaresults}, it can be seen that our method still achieves the best quantiative performance among all methods,
%better performance than all the compared methods,
%even though the spectral characteristic of \textit{Pavia University} is completely  different from that of ICVL,
which strongly demonstrates the advantage and generalization ability of our method.
Besides, the advantage of our method is further demonstrated by the visual comparison in Fig. \ref{fig:paviau}, where it can be seen that it is hard for the compared methods to completely remove the mixture noise
%in the synthetic remote sensing HS image is hard to remove completely by the compared methods.
By contrast, our method produces the denoised image with clearer and sharper textures. %which further demonstrates its advantage.

\subsubsection{Real noisy data}
Fig. \ref{fig:urban} provides the visual comparisons of denoised HS images by  different methods on Urban, a real-world noisy HS image, %noise the real-world noisy data (i.e., Urban).
where we can see that most of the compared methods fail to remove the unknown noise completely.
%completely remove such a complex real noise.
By contrast, our method successfully tackles this unknown noise and produces a clearer and visually pleasing image.
%comparable results in both detail preservation and noise removing, consistently demonstrating the generalization ability of our method.

\subsection{Ablation Study}
%\subsubsection{The effect of \textcolor{red}{filter} diversity.}
\label{subsubsec:abexp-filterrank}
We also directly and comprehensively compared the various combinations of low-dimensional convolutional kernels illustrated in Fig. \ref{fig:low-rank-filters}. %\st{feature extraction modules} over ICVL \cite{Arad2016Sparse} dataset with the common Gaussian noise level $\sigma=30$
%in Table \ref{tab:differentfilters}.
%\st{Especially, the 1-D low-dimensional filters and 1-D + 2-D filters without rank-enhancing, which extract HS features in the sequential way, are also included for investigating the effect of filter diversity.}
For fair comparisons, we built various denoising methods by only replacing the 1-D convolutional kernels of our method with the variants and retaining  all the other settings (e.g., connections, aggregation, etc.). %these HS image denoising frameworks via embedding each module into the U-Net in the same manner.
Besides, we also provided the results of 3-D convolution for reference.
%The framework constructed with the original 3-D filters is utilized as the baseline.
As listed in Table \ref{tab:differentfilters}, it can be seen that %\st{observed that all the feature extraction modules can achieve comparable performance.}
compared with the 3-D convolution, all the combinations of low-dimensional convolution  show their advantages on either quantitative performance or network compactness (\#Params) and complexity (\#FLOPs). %  feature extraction modules show their advantages in reducing parameters, \#FLOPs, and running time of the networks.
Generally,  a higher upper bound of the rank produces better  reconstruction quality, which is consistent with our theoretical analysis.
%More importantly,
%with the increasing of filter diversity, i.e., the upper limits of ranks of corresponding filter matrices are increased from $M$ to $3M$, the performance of HS image denoising is gradually improved.
Particularly, our Re-ConvSet equipped with the second fewest number of network parameters has the highest rank upper bound during the feature embedding process and thus achieves the best quantitative performance,
%with rank-enhanced 1-D low-dimensional filters achieve best performance among these feature extraction modules,
convincingly demonstrating its superiority and the importance of filter diversity in designing feature extraction module.

\section{Conclusion and Future Work}
\label{sec:con}

In this paper, we first proposed  Re-ConvSet, an efficient and effective module for extracting  high-dimensional spatial-spectral information of HS images. Specifically, %we quantitatively estimated the filter diversity via exploring the rank of unfolded filter matrix.
based on the theoretical analysis that improving the rank of the matrix formed by unfolded convolutional filters can promote feature diversity, %high-diversity and high-rank,
we designed Re-ConvSet by separately performing 1-D convolution along the three dimensions of HS images side-by-side and then aggregating the output, thus making it not only well capture the high-dimensional spatial-spectral information of HS images but also reduce the computational complexity.
Furthermore, we built our HS image denoising method by incorporating Re-ConvSet into the widely-used U-Net architecture. We conducted extensive experiments on both synthetic and real noisy HS images and demonstrated the significant superiority of such a concise method over state-of-the-art methods both quantitatively and visually.

In the future, we will explore the potential of the proposed Re-ConvSet in other kinds of HS image processing tasks, e.g., HS image spatial resolution, classification, segmentation, etc.

% or


%\appendices
%% you can choose not to have a title for an appendix
%% if you want by leaving the argument blank
%\section{}


\ifCLASSOPTIONcaptionsoff
  \newpage
\fi



%\bibliographystyle{IEEEtran}
%\begin{thebibliography}{1}
%\end{thebibliography}
\balance
\bibliographystyle{IEEEtran}
\bibliography{HSID}


%\end{IEEEbiography}
%% insert where needed to balance the two columns on the last page with
%%\newpage


%\vfill


\end{document}